\documentclass[trackchanges]{aastex701}

\received{March 25, 2026}
\revised{June 20, 2026}

\begin{document}

\title{Searching for Fast Astronomical Transients in Archival Photographic Plates}

\author{Ivo Busko}
\email{ivobusko@gmail.com}

\collaboration{all}{\textit{\footnotesize{Independent Researcher, Retired Developer at NASA/AURA/STScI}}}

\begin{abstract}

Fast astronomical transients were observed  by the VASCO Project 
\citep{Villarroel2020} in photographic sky surveys conducted in the 1950s.
Those searches analyzed the Palomar Observatory Sky Survey (POSS-I and POSS-II) digitized plates. 
In this article, we present a preliminary report on a similar but independent 
search using archival plates taken at the Hamburg Observatory with the Großer Schmidtspiegel 
1.2-m Schmidt camera, also from the mid-1950s. These plates were digitized by the  APPLAUSE 
Archive, which provides both images and tables of detected objects. 
By analyzing pairs of plates taken in rapid sequence (about 30 minutes apart) of the same sky regions, we find 
evidence of transients similar to those previously reported by the VASCO Project for POSS plates.
While the analysis is ongoing, one notable result is that our findings independently confirm that these transients 
seem to exhibit systematically 
narrow full width at half maximum (FWHM) compared to stellar point spread functions. If confirmed, this would
provide further support for their interpretation as sub-second optical flashes, as proposed by the VASCO team.
\end{abstract}


\section{Introduction} \label{sec:intro}

The Vanishing \& Appearing Sources during a Century of Observations (VASCO) project \citep{Villarroel2020} 
detected a number of fast transient  events using celestial observations from photographic 
plates.
All these observations are based on the Palomar Observatory Sky Surveys conducted 
in the 1950s. The favored interpretation of these events is that, absent reasonable astronomical
explanations for their very odd appearance and behavior, they nevertheless are consistent with 
the hypothesis that orbiting, or passing, artificial objects near Earth could create sub-second
glints from reflected Sun light \citep{villaroel2025aligned}. Given that these observations were
obtained at a time that predates our first artificial satellite, it becomes especially 
important to confirm the VASCO detections with independent experiments.

We looked for other archives of astronomical photographic plates that could serve as a 
basis for an independent search. Such archives must provide plate data in digitized form, 
accessible via online tools, and contain a significant number of plates obtained in the pre-Sputnik
era. We found that the digitized plates and ancillary data provided by the Archives of Photographic PLates for Astronomical USE (APPLAUSE) \footnote{\url{ https://www.plate-archive.org/cms/home/}} are very suitable for the purposes of this project. Aside from the digitized scans, it provides
tables of detected objects in each plate, obtained by running the SExtractor software 
\citep{1996A&AS..117..393B} 

The most important difference between the methodology used in this work, and the one used by the 
VASCO team \citep{Solano2022}, is that we use no external catalogs as a source of reference positions 
for objects recorded in the plates. We work only with plate pairs that share a common FOV on the sky, and
are separated by a short time interval (ideally taken in the same night.). The VASCO search starts by 
searching on any given plate, objects taken from a number of astronomical catalogs, and then works with
objects that fail identification. This approach has the advantage of being able to access almost any plate
at any region on the sky, at any time. Our approach gives access to a much more limited set of plates. Since
our purpose, at least initially, is not to study the phenomenon in a comprehensive way, but just to 
confirm/follow up
the work done by VASCO, use of a more limited data set doesn't pose a problem. And, by using a 
different methodology, any confirmation that comes from this study can potentially add even more 
significance to the discoveries.

\section{Data} \label{sec:data}

The APPLAUSE archive contains about 98,000 scanned plates, and about 4.5 billion extracted sources.
We started by looking, among the 27 plate archives/collections from Hamburg, Bamberg, Potsdam, 
Tautenburg, Tartu and Vatican Observatories, which collection / telescope would make a suitable target for
searching. Given that the VASCO Project focussed on plates taken by a large Schmidt camera, we looked into the holdings associated with the Hamburger Sternwarte Großer Schmidtspiegel 1.2-m camera. It has in principle very similar optical
and mechanical characteristics as the Samuel Oschin telescope at Palomar Observatory. Thus we should have similar plate scales, plate sizes, resolutions, and limiting magnitudes.

In this paper, we present a short description of the methodology, and report the preliminary findings of a search of transients in the plates produced by the Großer Schmidtspiegel camera. 

About 900 digitized plates can be found for that telescope, covering dates from 1954 to 1957 (later dates are avoided because we want skies free of human contaminants - Sputnik was launched in Oct 4, 1957). 
The plates cover typically FOVs about 2 X 2 deg, and were scanned at a resolution of 0.91 arcsec/pixel. 
The resulting digital files are about 10,000 x 10,000 pixels in size. A large majority of the plates from that 
plate set were scanned twice by the APPLAUSE team, rotating 
the plate by 90 deg over the scanner bed in between scans (generating X and Y scans). Each one of these 
scans were separately analyzed by SExtractor, producing two different lists of sources for each plate. 

Searching the archive catalog for at least three plates that share a common footprint on the sky (FOVs overlapping 50\% or more), share the same date, and were taken in quick succession, we found 41 plates that fulfill these requirements. This initial 
data set is used here primarily to develop and test possible search procedures. Planned work will augment the number of plates to be analyzed, by including plates sharing a common FOV but taken on separate nights, as well as data sets with only two plates..  

The basic data set for each plate includes two plate scans, and two source tables: one for raw scan quantities, and another for calibrated quantities. Both the raw and calibrated tables contain data from both X and Y scans.
They can be told apart in the table via the unique source ID and scan number associated with each object 
detected. 

\section{Analysis} \label{sec:analysis}

A very time and resource-consuming part of this kind of analysis is to just examine the entire plate scan with source-identification software. Also, astrometric solutions are mandatory for the plate scans to be usable in any form. By running SExtractor and astrometry solving software  themselves, and providing us with excellent tables and astrometry solutions with a lot of information on each and every single detected feature on the photographic images, the APPLAUSE archive helped significantly with the feasibility of this project.

The analysis steps can be summarized this way:

\begin{itemize}
    \item look for suitable plates in the database that fulfill certain criteria, such as:
    \begin{itemize}
        \item plates taken before October 1957;
        \item plates from the desired telescope;
        \item plates taken in the same night;
        \item plate FOVs overlap by a significant area;
        \item at least three plates within these conditions.
    \end{itemize}
    \item all data for a sequence of such plates (scans, source tables) is downloaded from APPLAUSE. A
    sequence is formed by all plates with common FOV and taken on the same night;
    \item a sequence is broken down into pairs of consecutive exposures, and, for each one of those pairs:
    \begin{itemize}
    \item cross-match the first plate's source table, with the second plate's source table. Objects that 
    do not have a counterpart in the second plate - these are the candidates for (vanishing) transients;
    \item fit FWHM and other parameters (from model Gaussians, from empirical radial profiles, from shape analysis functions) onto each one of these non-matched sources;
    \item compare these parameters with the same parameters derived from sources {\it with} a counterpart;
    \item display results from the few promising candidates that result from the statistical comparison, 
    and vet them visually, and against the USNO \citep{USNO} and Gaia \citep{Gaia_DR3} catalogs..
    \end{itemize}
\end{itemize}

The requirement that at least three plates be present in each sequence will be relaxed in a forthcoming iteration: because the core idea of the method is to compare a given plate with another plate of the same field, minimum 
sequence size is 2. There are, for this telescope, about 60 additional pairs of plates (taken on the same night) that fulfill the requirements.

The analysis steps were implemented in a series of Jupyter notebooks \citep{ivo_busko_2026}. These notebooks can be run 
individually, and also in batch mode (pipeline) over one entire sequence of plates, or a series of sequences. 
Manual running of a particular notebook over a selected subset of the data is sometimes necessary to 
fine-tune certain analysis parameters for certain plates or sequences.

The entire process starts with extracting a suitable ``master'' table from APPLAUSE that contains all 
necessary data to control what will be downloaded from the archive. Notebook {\it footprint\_analysis.ipynb} 
contains, at the top, an example of a SQL query that can be used to extract the master table. 
The {\it footprint\_analysis.ipynb} notebook reads the master table and looks for suitable plate pairs and sequences. Once found, the pipeline can be run over the desired sequences.

The main step in the pipeline is carried out by notebook {\it find\_mismatches.ipynb}, which cross-matches
sources for the first plate in the pair with sources for the second plate, looking for matches in celestial 
coordinates (within 5 arcsec). 

At the very beginning of this step, the code apply filters/thresholds to several 
SExtractor-generated parameters, such as extraction flags, elongation, position on plate, peak flux, model prediction. 
This removes the most blatant artifacts and bad quality star images from subsequent processing. 

At the cross-match step, we exploit the fact there are two scans for each plate, made in directions perpendicular to each other. Before looking for the mismatches between first and second plates, as described above, the code 
uses the same cross-matching procedure, but looking for mismatches between the X and Y scans {\it of the first 
plate}. That way, any artifact introduced by the scanning process, such as dirt on the scanner bed, dirt intoduced
by manipulation, and dirt 
that was somehow dislodged from, or acquired by,  the plate when it gets rotated, can be filtered out by removing detections
that appear on the X scan but do not have a counterpart on the Y scan. That significantly reduces the number of artifacts that get included in the output of this step.

One should note that the Y scan image itself is not needed for this analysis, but just its associated
source table. In practice, sources from both scans are included in one single table as delivered by 
APPLAUSE. We do not download the corresponding Y image file.

The PSF analysis step generates a number of diagnostics that help to understand the image characteristics, 
and hopefully could be of use in segregating real transients, from artifacts. First, it fits Gaussians to a subset
of stars, that is, objects that {\it do have} a match in the second plate. Next, it fits Gaussians to all non-matched
objects. Results are plotted and can be used to help figure out which parameters to use in filtering for 
candidates.

\begin{figure*}[ht!]
\plotone{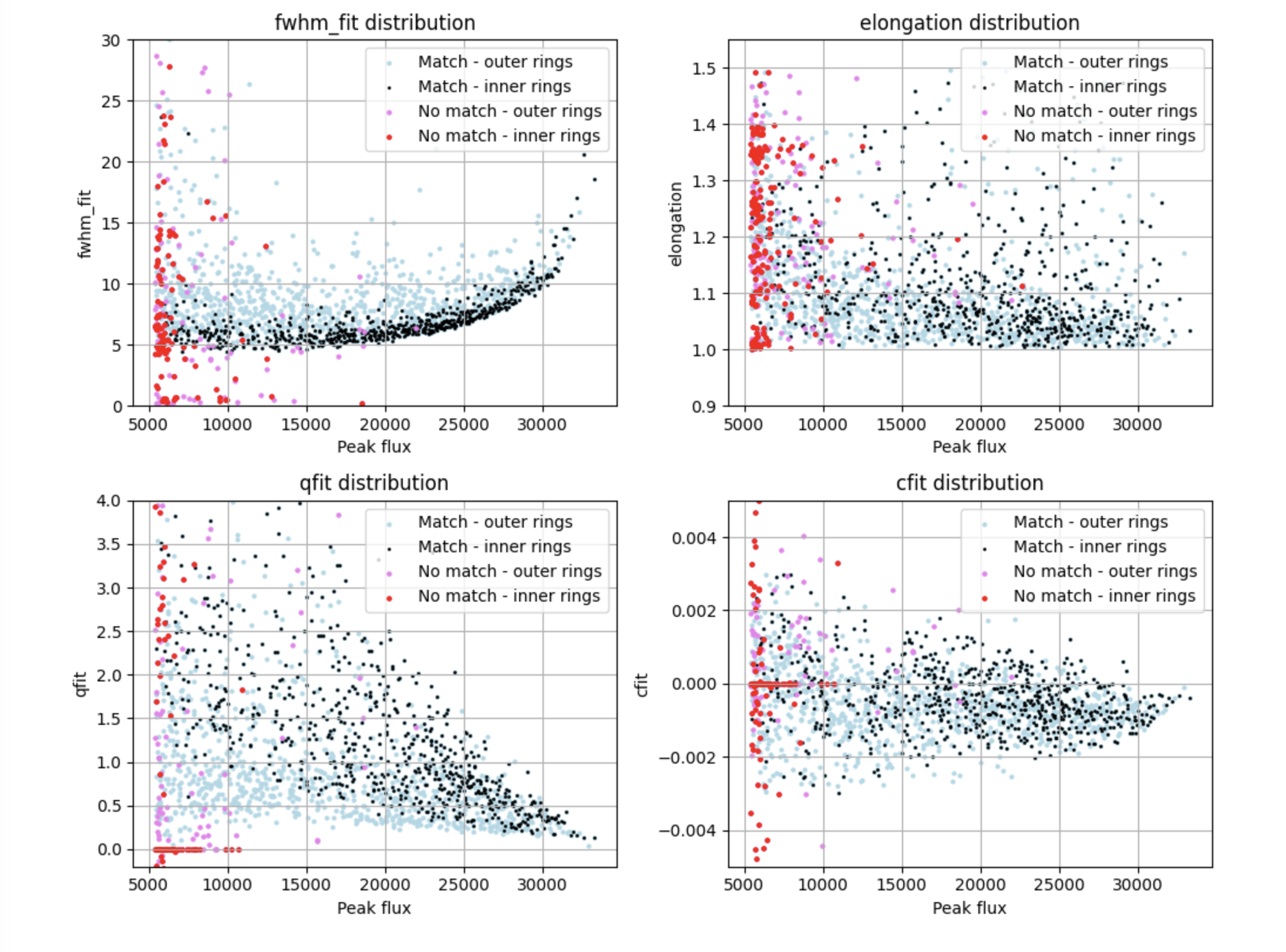}
\caption{One of the output diagnostics from the PSF analysis, for plate pair 9319-9320}
\label{fig:psf}
\end{figure*}

One could argue that a Gaussian is not a perfect model for a star image on a photographic plate, based on
the fact that photographic density has a logarithmic dependency on exposure (a better model would be
perhaps a quadratic). However, in practice the approach seems to work satisfactorily; the diagnostics 
generated by the fit procedure (Fig. \ref{fig:psf}) show how much the actual data departs from the 
Gaussian model.  The departure is consistent with the quadratic model. In particular, FWHM tends to 
be underestimated.

A more important factor is the position on the plate, since FWHM is clearly dependent on radial distance
from plate center, on some plates. Certainly a telescope focus issue; it should be accounted for when vetting
candidates.

The analysis step also computes radial profiles for each non-matched object, and for a number of stars 
in its neighborhood ($\sim6$ arcmin), and with peak flux within 0.1 mag of its peak flux. It also computes
some shape parameters for a fixed isophote, as an attempt to characterize the object's shape (this is
still being worked out).

A filtering step is applied before the output of the analysis step, removing anything that has FWHM 
and peak flux outside limits specified by user-settable parameters. 

Finally, the display notebook creates visualizations for all surviving non-matched sources, after applying 
additional filtering based on shape parameters. Due to false positives in the SExtractor table for the
second plate (when the table doesn't list an entry for a visible but faint object),
an additional filter based on aperture photometry is used to reject these (few) false positives. 
A cross-match against the USO and Gaia DR3 catalogs is also done at this step, to reject real stars
that somehow managed to survive the analysis to this point.

The pipeline includes additional steps to collate and summarize results.

The code in the notebooks exploits the fact laptop computers can provide a significant degree of 
parallelization (10 processors on a MacBook M1). That, on top of significant gains in performance enabled
by vectorized operations in the {\it numpy} library \citep{harris2020array}, rendered this project feasible at
no cost on a personal laptop computer.

\begin{figure*}[ht!]
\plotone{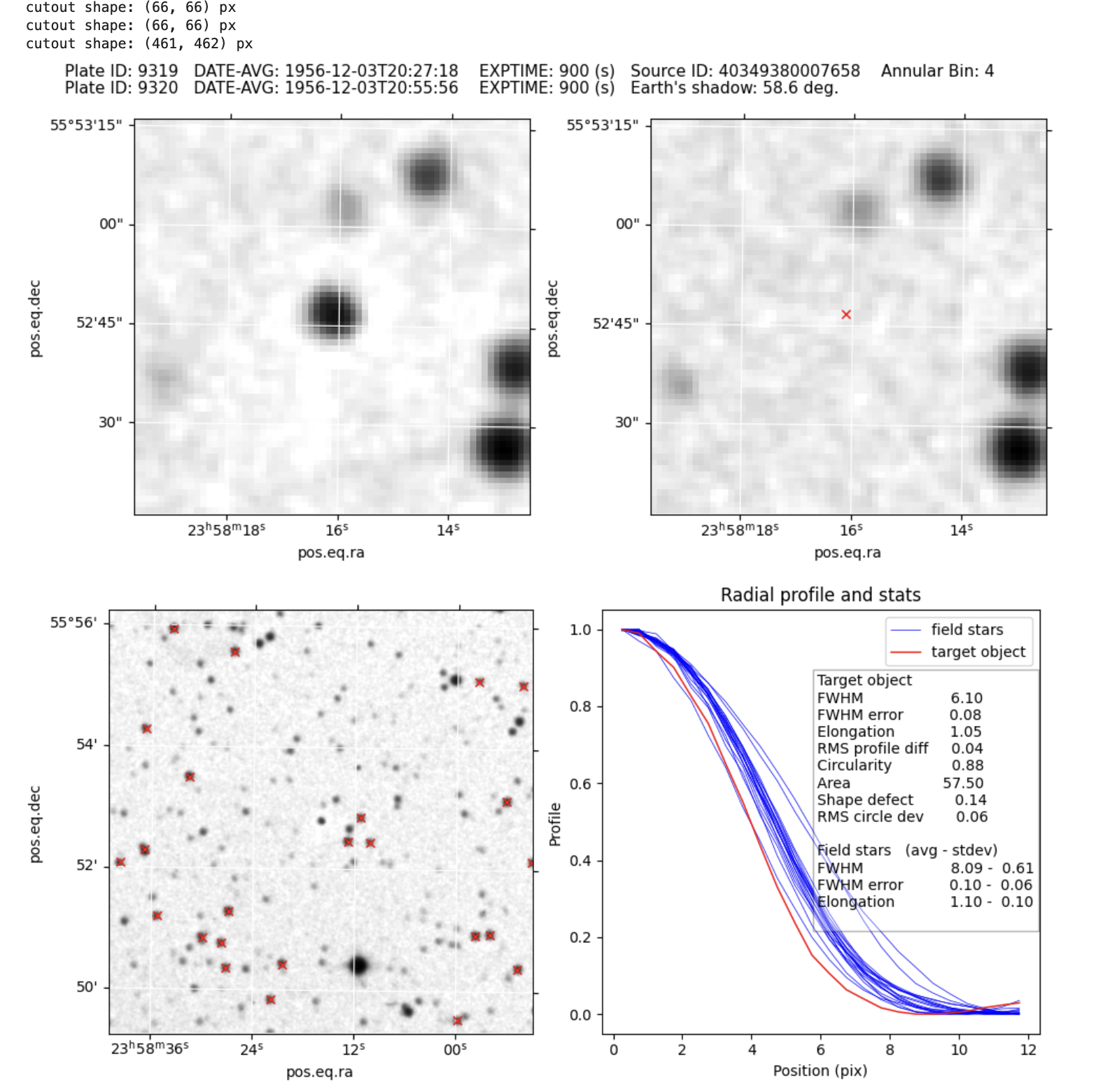}
\caption{Output of the 'display' pipeline step for one particular transient candidate. {\bf Top left:}
transient on the first plate; {\bf Top right:} position of transient on the second plate (about 30 min.
later); {\bf Bottom left:} 6 arcmin neighborhood around transient; the marked stars lie within 0.1 mag
of the transient itself; {\bf Bottom right:} normalized radial profiles, and stats.}
\label{fig:display}
\end{figure*}

\section{Results} \label{sec:results}

Results to date are summarized in the {\it results.ipynb} notebook in the software repository \citep{ivo_busko_2026}. The 41 plates analyzed so far yielded 70 candidates; visual vetting
dropped that to 35 ``good candidates''.  

Some preliminary results related to the analysis process itself were found. In particular, we found that
the FWHM by itself, as derived by Gaussian fitting,  is not a reliable discriminator against artifacts: it can 
depend on the position on the plate, and can  be biased by the non-Gaussian nature of the light distribution
of each detected source. The radial brightness profile seems to be more useful, by comparing
the source profile with an average profile of stars with similar magnitude in the neighborhood.  

Shape parameters are still being evaluated as a discrimination tool, but we likely need a larger 
sample before trying to use them as discrimination criteria. 

While the final interpretation of the morphological properties of the detected transients is still 
under analysis, one noteworthy result should be highlighted regarding their FWHM. 
Specifically, the events exhibit systematically smaller FWHM compared to stellar images, 
consistent with their interpretation as extremely short-duration flashes (see Fig. \ref{fig:display}). 
As discussed by \citep{V2025a}, unresolved flashes lasting less than a second naturally appear sharper and more circular than stellar images, particularly on long-exposure plates where stars are significantly 
blurred by seeing and tracking errors. Such profiles are therefore 
an expected observational signature of sub-second optical flashes, further reinforcing the transient interpretation.

\section{Conclusions} \label{sec:conclusions}

This paper describes the basics of an ongoing experiment designed to independently verify
the findings of the VASCO project. In particular, it addresses the existence and properties 
of fast celestial transients recorded in archival photographic plates taken decades ago. 
While such transients are difficult to reconcile within a conventional astronomical framework, 
they are consistent with sub-second optical glints produced by 
sunlight reflecting from flat surfaces on rotating objects transiting above Earth’s atmosphere. 
Given that these observations predate the space era, establishing a robust observational basis 
for the reality and behavior of these events is of clear importance.

Using astronomical plates taken in the 1950s with the Großer Schmidtspiegel camera at Hamburger Sternwarte, and digitized within the  APPLAUSE Archive, we obtain initial results that seem to 
independently confirm the presence of such transients. While the work is still ongoing, one 
notable result is the emerging consistency with previous findings from the VASCO project, particularly 
regarding the expected observational signatures of sub-second optical flashes in astronomical, 
long-exposure photographic plates. 

So far, only a small fraction of the available plates has been analyzed. Future work will expand 
the dataset to include additional plates from the same telescope, as well as material from other 
instruments within the APPLAUSE Archive. The goal is to assemble all events into a database, 
and then search for possible alignments, or other correlations, in between different events in the 
database. Ultimately, cross-correlation with transients already identified by the VASCO project 
will be essential for advancing our understanding of these phenomena.

\begin{acknowledgments}

Funding for APPLAUSE has been provided by DFG (German Research Foundation, Grant ), 
Leibniz Institute for Astrophysics Potsdam (AIP), Dr. Remeis Sternwarte Bamberg (University 
Nuernberg/Erlangen), the Hamburger Sternwarte (University of Hamburg) and Tartu Observatory.

\end{acknowledgments}

\software{plateanalysis \citep{ivo_busko_2026},
          astropy \citep{astropy:2013, astropy:2018, astropy:2022},
          Source Extractor \citep{1996A&AS..117..393B},
          numpy \citep{harris2020array}
          }
\bibliographystyle{aasjournalv7}

\bibliography{paper01}{}

\end{document}